\documentclass[a4paper,12pt]{article}

\usepackage{amsmath,amssymb}
\usepackage{cite}

\begin{document}

\author{Nikolai A. Kudryashov \footnote{E-mail: Kudryashov@mephi.ru}  and  Mikhail B. Soukharev\footnote{E-mail: Soukharev@mephi.ru}}

\title{Comment on: ''New exact solutions of the (3+1)-dimensional Burgers System''
[Phys. Lett. A 373 (2009) 181]}

\date{Department of Applied Mathematics, Moscow Engineering and Physics Institute, Kashirskoe sh. 31, Moscow 115409, Russia}

\maketitle

\begin{abstract}

We demonstrate that all exact solutions of the Riccati equation by  Dai and Wang [C.-Q. Dai, Y.-Y. Wang, Phys. Lett. A 373 (2009) 181--187] are not new and cannot be new because the general solution of this equation was obtained more than one century ago. Moreover we show that some ''new solutions'' by Dai and Wang of the Riccati equation do not satisfy this equation.
We also illustrate that the authors did not obtain any new solutions for solution of the (3+1)-dimensional Burgers system.

\end{abstract}



\section{On ''new'' solutions of the Riccati equation by Dai and Wang}

Dai and Wang \cite{Dai_2009} looked for exact solutions of the Riccati equation
\begin{equation}
  \frac{d\varphi(\xi)}{d\xi}=l_0+\varphi^2(\xi) \label{Riccati 1},
\end{equation}
where $l_0$ is constant. In their letter these authors say: "we firstly use the the Exp---function method \cite{He_2006} to seek new exact solutions of the Riccati equation \eqref{Riccati 1}". Firstly Dai and Wang are wrong here because they are not first who applied the Exp-function method to find "new solutions" of the Ricatti equation. Zhang  was first in \cite{Zhang_2008} who used the Exp - function method  to find "new generalized solitonary solutions of Riccati equation". Criticism of the paper by Zhang \cite{Zhang_2008} was given in our recent work \cite{Kudr_2009a}.

Secondly Dai and Wang claim: ''we obtain some new exact
solutions of the Riccati equation''. The Riccati equation \eqref{Riccati 1} has been
studied during several centuries, therefore statement by Dai and Wang may cause a sensation.
Unfortunately there is no sensation because this statement is wrong as well.

Let us show that some ''solutions'' by Dai and Wang are not new and some of them do not satisfy the Riccati equation.
"Introducing a complex variable" $\eta=k\xi+\xi_0$ Dai and Wang rewrite equation
\eqref{Riccati 1} in the form
\begin{equation}
  k\varphi'-l_0-\varphi^2=0,\label{RiccatiODE}
\end{equation}
where $k$ is a constant, $\xi_0$ is an arbitrary constant. It is well known (see any
textbook on differential equations) that this equation has the general solution in the form
\begin{equation}
  \varphi(\eta)=-\sqrt{-l_0}\tanh\left(\frac{\sqrt{-l_0}}{k}\eta\right),\quad
  \eta=k\xi+\xi_0. \label{general solution}
\end{equation}
This solution depends on one arbitrary constant $\xi_0$. In the limit $k\to\mp0$ this
formula degenerates to the constant solution
\begin{equation}
  \varphi(\eta)=\pm\sqrt{-l_0}. \label{constant solution}
\end{equation}
There is no any other solution of equation \eqref{RiccatiODE} besides \eqref{general
solution} and \eqref{constant solution}.

Using the Exp-function method Dai and Wang found \emph{five exact solutions} of the Riccati
equation \eqref{RiccatiODE}. These solutions are as follows:
\begin{equation}
  \varphi_1=\frac{-\sqrt{-l_0}b_1e^\eta+a_{-1}e^{-\eta}}
  {b_1e^\eta+\frac{a_{-1}}{\sqrt{-l_0}}e^{-\eta}},\quad \eta=\sqrt{-l_0}\xi+\xi_0,
  \label{Rsolution1}
\end{equation}

\begin{equation}
  \varphi_2=\frac{-i\sqrt{l_0}b_1e^\eta+a_{-1}e^{-\eta}}
  {b_1e^\eta-i\frac{a_{-1}}{\sqrt{l_0}}e^{-\eta}},\quad \eta=i\sqrt{l_0}\xi+\xi_0,
  \label{Rsolution2}
\end{equation}

\begin{equation}
  \varphi_3=\frac{-\sqrt{-l_0}b_1e^\eta+a_0-\sqrt{-l_0}b_{-1}e^{-\eta}}
  {\frac{a_0^2+l_0b_0^2}{4l_0b_{-1}}e^\eta+b_0+b_{-1}e^{-\eta}},\quad
  \eta=2\sqrt{-l_0}\xi+\xi_0, \label{Rsolution3}
\end{equation}

\begin{equation}
  \begin{gathered}
    \varphi_4=\frac{\sqrt{-l_0}b_2e^{2\eta}+a_1e^\eta+a_0-
    \frac{\sqrt{-l_0}}{l_0}b_{-1}e^{-\eta}}{b_2e^{2\eta}+b_1e^\eta+
    \frac{a_1^2+l_0b_1^2+2\sqrt{-l_0}a_0b_2}{2l_0b_2}+
    \frac{\left(a_1+\sqrt{-l_0}b_1\right)\left(4a_0b_2-
    \sqrt{-l_0}a_1^2-l_0\sqrt{-l_0}b_1^2\right)}{8l_0^2b_2^2}e^{-\eta}},\\
    \eta=2\sqrt{-l_0}\xi+\xi_0, \label{Rsolution4}
  \end{gathered}
\end{equation}

\begin{equation}
  \varphi_5=\frac{\sqrt{-l_0}b_2e^{2\eta}-\sqrt{-l_0}b_1e^\eta-\frac{a_{-1}b_2}{b_1}
  +a_{-1}e^{-\eta}}{b_2\exp(2\eta)+b_1e^\eta+\frac{\sqrt{-l_0}a_{-1}b_2}{l_0b_1}+
  \frac{\sqrt{-l_0}}{l_0}e^{-\eta}},\quad \eta=-2\sqrt{-l_0}\xi+\xi_0.
  \label{Rsolution5}
\end{equation}
Solutions \eqref{Rsolution1}--\eqref{Rsolution5} correspond to formulae (17), (20), (23),
(27) and (28) in the work \cite{Dai_2009}.

Note that the Riccati equation \eqref{RiccatiODE} is the first-order differential equation and all solutions of this equation can have \emph{only one arbitrary constant} \cite{Kudr_2009b, Kudr_2009c, Kudr_2005, Kudr_2008}.  Therefore formulae \eqref{Rsolution1}--\eqref{Rsolution5}
can contain only one arbitrary constant. But Dai and Wang believe that there are more
arbitrary constants in each expression \eqref{Rsolution1}--\eqref{Rsolution5}. Namely,
they claim that there are: two arbitrary constants $a_{-1}$ and $b_1$ in $\varphi_1$ and
$\varphi_2$, three arbitrary constants $a_0$, $b_0$, and $b_{-1}$ in $\varphi_3$, four
arbitrary constants $a_0$, $a_1$, $b_1$, and $b_2$ in $\varphi_4$, three arbitrary
constants $a_{-1}$, $b_1$, and $b_2$ in $\varphi_5$.

Let us demonstrate that this is not the case because it is not possible never.
Rewriting expression \eqref{Rsolution1} we have
\begin{equation}
  \begin{gathered}
    \varphi_1=\frac{-\sqrt{-l_0}b_1e^\eta+a_{-1}e^{-\eta}}
    {b_1e^\eta+\frac{a_{-1}}{\sqrt{-l_0}}e^{-\eta}}=
    -\sqrt{l_0}\frac{1-\frac{a_{-1}}{\sqrt{-l_0}b_1}e^{2\eta}}
    {1+\frac{a_{-1}}{\sqrt{-l_0}b_1}e^{2\eta}}={}\\
    {}=-\sqrt{-l_0}\tanh\left(\eta-\frac12\log\frac{a_{-1}}{\sqrt{-l_0}b_1}\right).
  \end{gathered}
\end{equation}
Due to $\eta=\sqrt{-l_0}\xi+\xi_0$ we obtain that there is only one arbitrary constant
$\xi_0-\frac12\log\frac{a_{-1}}{\sqrt{-l_0}b_1}$ in the argument of tanh. Therefore
solution $\varphi_1$ by Dai and Wang coincides with known solution \eqref{general
solution}.

We can see that solution $\varphi_2$ by Dai and Wang is equal to $\varphi_1$ because
$\sqrt{-l_0}=i\sqrt{l_0}$. Therefore $\varphi_2$ coincides with solution \eqref{general
solution}.

Substituting $\varphi_3$ in Eq.\eqref{RiccatiODE} we do not get zero. Therefore expression
$\varphi_3$ does not satisfy Eq.\eqref{RiccatiODE} and this expression is not a solution of the Riccati equation \eqref{RiccatiODE} if $a_0$, $b_0$,
$b_{-1}$ are arbitrary constants. Function $\varphi_3$ can be solution of the
Riccati equation \eqref{RiccatiODE} only if we assume additional constraints
\begin{equation}
  a_0^2+b_0^2l_0=0,\quad b_{-1}\left(b_0l_0-a_0\sqrt{-l_0}\right)=0
\end{equation}
However in this case expression $\varphi_3$ is the trivial solution \eqref{constant
solution}.

Substituting $\varphi_4$ in equation \eqref{RiccatiODE} we do not obtain zero as well. So
function $\varphi_4$ is not a solution of the Riccati equation \eqref{RiccatiODE} if
$a_0$, $a_1$, $b_1$, and $b_2$ are arbitrary constants. Expression $\varphi_4$
is a solution of the Riccati equation \eqref{RiccatiODE} only if we take additional constraints
into account \begin{equation}
  a_1=-\sqrt{-l_0}b_1,\quad b_1b_2=0,\quad a_0b_2=0,
\end{equation}
or
\begin{equation}
  a_1=\sqrt{-l_0}b_1,\quad a_0b_2=0,\quad a_0b_1=0.
\end{equation}
In these cases expression $\varphi_4$ is the trivial solution
\eqref{constant solution} again.

Expression \eqref{Rsolution5} can be presented as the following
\begin{equation}
  \begin{gathered}
    \varphi_5=\frac{\sqrt{-l_0}b_2e^{2\eta}-\sqrt{-l_0}b_1e^\eta-\frac{a_{-1}b_2}{b_1}+
    a_{-1}e^{-\eta}}{b_2e^{2\eta}+b_1e^\eta+\frac{\sqrt{-l_0}a_{-1}b_2}{l_0b_1}+
    \frac{\sqrt{-l_0}}{l_0}a_{-1}e^{-\eta}}={}\\
    {}=\sqrt{-l_0}\frac{\left(b_2e^\eta-b_1\right)
    \left(e^\eta-\frac{a_{-1}}{b_1\sqrt{-l_0}}e^{-\eta}\right)}
    {\left(b_2e^\eta+b_1\right)
    \left(e^\eta-\frac{a_{-1}}{b_1\sqrt{-l_0}}e^{-\eta}\right)}={}\\
    {}=\sqrt{-l_0}\frac{\left(b_2e^\eta-b_1\right)}{\left(b_2e^\eta+b_1\right)}=
    \sqrt{-l_0}\tanh\left(\frac\eta2+\frac12\log\frac{b_2}{b_1}\right).
  \end{gathered}
\end{equation}
Due to $\eta=-2\sqrt{-l_0}\xi+\xi_0$ we obtain that there is only one arbitrary constant
$\frac{\xi_0}{2}+\frac12\log\frac{b_2}{b_1}$ in argument of tanh. Therefore solution
$\varphi_5$ by Dai and Wang coincides with known solution \eqref{general solution}.

Thus we have proved that Dai and Wang did not find new solutions of the Riccati equation
\eqref{RiccatiODE}. Moreover, functions $\varphi_3$ and $\varphi_4$ in general case are
not solutions of the Riccati equation \eqref{RiccatiODE}. What is more we state that nobody can not obtain new exact solutions of the Riccati equation \eqref{RiccatiODE}. The statement by Dai and Wang in \cite{Dai_2009} on the Riccati equation is wrong.

\section{On ''new'' solutions of the (3+1)-dimensional Burgers system by Dai and Wang}

Dai and Wang  have considered the (3+1)-dimensional Burgers system in \cite{Dai_2009} as  well
\begin{equation}\begin{gathered}
  u_t-2uu_y-2vu_x-2wu_z-u_{xx}-u_{yy}-u_{zz}=0,\\
  \\ u_x-v_y=0,\quad u_z-w_y=0.
  \label{Burgers system}
\end{gathered}\end{equation}
Authors \cite{Dai_2009} claim: "based on the Riccati equation and its new exact solutions, we find new and more general exact solutions with two arbitrary functions of the
(3+1)-dimensional Burgers system". In this section we show that this statement is wrong.
All solutions of the (3+1)-dimensional Burgers system by Dai and Wang are not new.

Authors \cite{Dai_2009} have obtained two formal solutions of the system of equations \eqref{Burgers system}
\begin{equation}
  u=-h\varphi(\xi),\quad
  v=\frac{p_t-p_{xx}-p_{zz}}{2p_x}-p_x\varphi(\xi),\quad
  w=-p_z\varphi(\xi)\label{f sol 1}
\end{equation}
and
\begin{equation}
  \begin{gathered}
    u=-\frac12h\varphi(\xi)+\frac12h\sqrt{l_0+\varphi^2(\xi)},\\
    v=\frac{p_t-p_{xx}-p_{zz}}{2p_x}-\frac12p_x\varphi(\xi)+
    \frac12p_x\sqrt{l_0+\varphi^2(\xi)},\\
    w=-\frac12p_z\varphi(\xi)+\frac12p_z\sqrt{l_0+\varphi^2(\xi)}.\label{f sol 2}
  \end{gathered}
\end{equation}
Here $\varphi(\xi)$ is solution of the Riccati equation \eqref{Riccati 1}. In formulae
\eqref{f sol 1}--\eqref{f sol 2} Dai and Wang take $\xi=p(x,z,t)+hy$, where $p(x,z,t)$ is
an arbitrary function, $h$ is an arbitrary constant.

Substituting expressions \eqref{Rsolution1}--\eqref{Rsolution5} to the formulae \eqref{f
sol 1}--\eqref{f sol 2} Dai and Wang obtained ten ''new solutions'' of the Burgers system
\eqref{Burgers system}. We proved in previous section  that expressions
\eqref{Rsolution1}--\eqref{Rsolution5} by Dai and Wang are equivalent to two solutions
\eqref{general solution} and \eqref{constant solution} of the Riccati equation. Therefore
Dai and Wang can get only four solutions of the Burgers system \eqref{Burgers
system}. Let us show that only two distinct solutions of \eqref{Burgers system} can be
obtained in such a way.

Taking solution \eqref{general solution} of the Riccati equation \eqref{Riccati 1} in the
form
\begin{equation}
  \varphi(\xi)=-\sqrt{-l_0}\tanh\left(\sqrt{-l_0}(p+hy)+\xi_0\right),\quad \xi=p+hy,
  \label{Riccati g sol}
\end{equation}
and substituting \eqref{Riccati g sol} into the formal solution \eqref{f sol 1} we have
\begin{equation}
  \begin{gathered}
    u=h\sqrt{-l_0}\tanh\left(\sqrt{-l_0}(p+hy)+\xi_0\right),\\
    v=\frac{p_t-p_{xx}-p_{zz}}{2p_x}+
    p_x\sqrt{-l_0}\tanh\left(\sqrt{-l_0}(p+hy)+\xi_0\right),\\
    w=p_z\sqrt{-l_0}\tanh\left(\sqrt{-l_0}(p+hy)+\xi_0\right). \label{true form 1}
  \end{gathered}
\end{equation}
Without loss of generality we can introduce new arbitrary function
$q(x,z,t)=\sqrt{-l_0}p(x,z,t)+\xi_0$ and new arbitrary constant $r=\sqrt{-l_0}h$. Then
expressions \eqref{true form 1} take the form
\begin{equation}\begin{gathered}
u=r\tanh(q+ry),\quad v=\frac{q_t-q_{xx}-q_{zz}}{2q_x}+q_x\tanh(q+ry),\\
\\
w=q_z\tanh(q+ry), \label{Li solution}
\end{gathered}\end{equation}

Now let us substitute the function \eqref{Riccati g sol} in the formal solution \eqref{f
sol 2}. After some transformations we have
\begin{equation}
  \begin{gathered}
    u=\frac12h\sqrt{-l_0}\tanh\left(\frac{\sqrt{-l_0}}{2}(p+hy)+
    \frac{\xi_0}{2}-\frac{i\pi}{4}\right),\\
    v=\frac{p_t-p_{xx}-p_{zz}}{2p_x}+\frac12p_x\sqrt{-l_0}
    \tanh\left(\frac{\sqrt{-l_0}}{2}(p+hy)+\frac{\xi_0}{2}-\frac{i\pi}{4}\right),\\
    w=\frac12p_z\sqrt{-l_0}\tanh\left(\frac{\sqrt{-l_0}}{2}(p+hy)+
    \frac{\xi_0}{2}-\frac{i\pi}{4}\right). \label{true form 2}
  \end{gathered}
\end{equation}
Introducing new arbitrary function
$q(x,z,t)=\frac{\sqrt{-l_0}}{2}p(x,z,t)+\frac{\xi_0}{2}-\frac{i\pi}{4}$ and new arbitrary
constant $r=\frac{\sqrt{-l_0}h}{2}$ we can rewrite expressions \eqref{true form 2} in the
form \eqref{Li solution}.

Therefore solutions \eqref{true form 1} and \eqref{true form 2} are the same. However solution
\eqref{Li solution} of the Burgers system \eqref{Burgers system} is not new. This solution was found by Li et al in work \cite{Li}.

Taking constant solution $\varphi(\xi)=\pm \sqrt{-l_0}$ of the Riccati equation
\eqref{Riccati 1} and substituting it in \eqref{f sol 1} we have
\begin{equation}
  u=\pm h\sqrt{-l_0},\quad v=\frac{p_t-p_{xx}-p_{zz}}{2p_x}\pm p_x\sqrt{-l_0},\quad w=\pm
  p_z\sqrt{-l_0}. \label{true form 3}
\end{equation}
Substituting the same function in \eqref{f sol 2} we obtain
\begin{equation}
  u=\pm\frac h2\sqrt{-l_0},\quad
  v=\frac{p_t-p_{xx}-p_{zz}}{2p_x}\pm\frac{p_x}{2}\sqrt{-l_0},\quad
  w=\pm\frac{p_z}{2}\sqrt{-l_0}. \label{true form 4}
\end{equation}
We can write solutions \eqref{true form 3}--\eqref{true form 4} in more general form
\begin{equation}
  u=r,\quad v=q,\quad w=s, \label{obvious solution}
\end{equation}
where $q=q(x,z,t)$ and $s=s(x,z,t)$ are arbitrary functions, $r$ is an arbitrary
constant. Therefore solutions \eqref{true form 3}--\eqref{true form 4} are the same.
Solution \eqref{obvious solution} of the system \eqref{Burgers system} is obvious and can
be obtained without any calculations.

Thus we see that Dai and Wang get only two distinct solutions \eqref{Li solution} and
\eqref{obvious solution} of the Burgers system \eqref{Burgers system}. Solution \eqref{Li
solution} is not new, solution \eqref{obvious solution} is the trivial solution. So the statement by Dai and Wang cited in the beginning of this section is wrong. Authors \cite{Dai_2009} did not obtain  new results for the Riccati equation \eqref{RiccatiODE} and for the Burgers system \eqref{Burgers system}. Some statements and some results by Dai and Wang are wrong.

{99}

\end{document}